\documentclass[apjl]{emulateapj}
\usepackage{graphicx,color}
\usepackage{amssymb,txfonts}
\usepackage{epsfig,natbib}
\usepackage{multirow}
\usepackage{rotating}
\newcommand{\fpath}{}
\newcommand{\red}{\color{red}}
\newcommand{\white}{\color{white}}

\DeclareMathAlphabet{\mathitbf}{OML}{cmm}{b}{it}
\DeclareMathAlphabet{\mathf}{OML}{cmm}{c}{sl}

\newcommand{\Bz}{\mathf B_z}

\newcommand{\Bazi}{\mathf B_\phi}
\newcommand{\Bzabs}{\mathit B_z}
\newcommand{\jz}{{\mathf j_z}}

\newcommand{\Enlff}{E_{\rm nlff}}
\newcommand{\Epot}{E_{\rm pot}}
\newcommand{\Efree}{E_{\rm free}}
\newcommand{\ie}{i.~e.}
\newcommand{\eg}{e.~g.}

\slugcomment{Accepted \today~for publication in \apj}

\begin{document}

\title{Force-free field modeling of twist and braiding-induced magnetic energy in an active-region corona}
\author{J.~K.~Thalmann\altaffilmark{1,2} and S.~K.~Tiwari\altaffilmark{2} and T.~Wiegelmann\altaffilmark{2}}
\email{julia.thalmann@uni-graz.at}
\altaffiltext{1}{Institute of Physics/IGAM, University of Graz, Universit\"atsplatz 5, 8010 Graz, Austria}
\altaffiltext{2}{Max Plank Institute for Solar System Research, Max-Planck-Str. 2, 37191 Katlenburg-Lindau, Germany}

\begin{abstract}
The theoretical concept that braided magnetic field lines in the solar corona may dissipate a sufficient amount of energy to account for the brightening observed in the active-region corona, has been substantiated by high-resolution observations only recently. From the analysis of coronal images obtained with the High Resolution Coronal Imager, first observational evidence of the braiding of magnetic field lines was reported by \cite{cir_gol_13} (hereafter CG13). We present nonlinear force-free reconstructions of the associated coronal magnetic field based on vector SDO/HMI magnetograms. We deliver estimates of the free magnetic energy associated to a braided coronal structure. Our model results suggest ($\sim$\,100 times) more free energy at the braiding site than analytically estimated by CG13, strengthening the possibility of the active-region corona being heated by field line braiding. We were able to assess the coronal free energy appropriately by using vector field measurements and attribute the lower energy estimate of CG13 to the underestimated (by a factor of 10) azimuthal field strength. We also quantify the increase of the overall twist of a flare-related flux rope which had been claimed by CG13. From our models we find that the overall twist of the flux rope increased by about half a turn within twelve minutes. Unlike another method, to which we compare our results to, we evaluate the winding of the flux rope's constituent field lines around each other purely based on their modeled coronal 3D field line geometry -- to our knowledge for the first time.
\end{abstract}


\section{Introduction}

The plasma of the solar corona is much hotter ($\gtrsim$\,$10^{6}$~K) than that of the photosphere ($\sim$\,6000~K). The mechanism that could result in such an extraordinarily heated solar corona is not yet distinctly understood. Several mechanisms, including nano-flares, Alfv\'en wave heating, MHD turbulence, heating by X-ray jets and bright points have been proposed but provide merely a partial solution to the coronal heating problem \citep[\eg,][]{wal_ire_03,asch_04,mci_dep_11,wed_scu_12,win_wal_13}. 

The density and temperature distributions in the active region (AR) and quiet-Sun corona are quite different. The plasma temperature in the AR corona is 8-20 $\times10^6$ K which is by a factor of 4-10 higher than that of the quiet-Sun corona \citep[\eg][]{zir_93}. The most widely believed phenomenon that accounts for the heating of the AR (magnetically-closed) corona is the braiding of (ensembles of) magnetic field lines \citep[which numbers their crossings; see \eg,][]{ber_asg_09}. It results in high temperatures either by Joule (ohmic) heating of currents induced by entangled magnetic field lines \citep{par_72} or by nano-flares occurring when neighboring, oppositely directed field lines reconfigure via magnetic reconnection \citep[][]{par_83,par_88,pri_02}. The latter is supported by theoretical models \citep[\eg][]{kli_06}. The former was recently investigated by \cite{bou_bin_13} who compared synthesized emission from a forward 3D MHD coronal model to actual coronal images. They were able to show that the field line braiding delivered an energy input required for the observed heating of the AR-corona by ohmic dissipation.

The braiding of magnetic field lines can be caused by random displacements of where magnetic field lines  are line-tied at a photospheric level (\ie~of their footpoints) or by vortical motions of the photospheric plasma, the latter resulting in the twisting (winding) of field lines about each other. Different MHD models on magnetic field line braiding have been developed and most of them support the former mechanism \citep[\eg,][and references therein]{gud_nor_05,rap_val_08,vba_asg_11}. 

The observational evidences of these processes have never been very clear, longing for high resolution observations. The development of recent space-based instruments, \eg, the Solar Optical Telescope (SOT) on board {\it Hinode} \citep[][]{kos_mat_07,tsu_08,sue_tsu_08}  and the Atmospheric and Imaging Assembly (AIA) on board {\it SDO} \citep{lem_tit_12}, and their delivery of high-resolution coronal images allow us to have a closer look to the mechanisms heating the coronal plasma. In particular, the data obtained from NASA's recently flown rocket carrying the High Resolution Coronal Imager \citep[Hi-C; see][and \cite{cir_gol_13}]{gol_cir_06} with a spatial resolution of $\sim$\,0.2$''$ (6 times that of AIA with $\sim$\,1.2$''$), have given an unique opportunity for a fresh look at the coronal heating mechanism. Using Hi-C data, \cite{cir_gol_13} (hereafter ``CG13'') claimed first observational evidence of the braiding mechanism to deliver the amount of energy required to heat the AR corona. However, a direct computation of the free magnetic energy stored in the AR loops was not possible due to the lack of direct coronal magnetic field measurements. In this work, we close this gap using a nonlinear force-free (NLFF) coronal magnetic field model to substantiate the estimated energy budget.

Another important aspect of the analysis of CG13 was the seemingly increasing twist of a magnetic structure during the rising phase of a small flare (which the 5-minute observation time of Hi-C covered). The twist of a magnetic structure is determined by the winding of the magnetic field lines around a central axis and is related to its helicity \citep[\eg][]{ber_99}. Attempts to estimate the twist of AR magnetic fields have been made based on the length of field lines and the force-free parameter where they are line-tied at a photospheric level \citep[\eg,][]{lea_can_03} but were suspected to underestimate an AR's global twist using a ``best-fit'' force-free parameter. \cite{lek_fan_05} suspected that only when applied to thin flux tubes this method may correctly recover the winding of the flux rope axis \citep[see also, \eg,][]{ino_shi_12}. On an AR scale, however, other guesses for a global value of the force-free parameter might be appropriate \citep[\eg,][]{tiw_ven_09}. Here, we try a novel approach to estimate the winding of a flux rope's constituent field lines in the corona by using their 3D geometry as inferred from the NLFF modeling. In this way we aim to verify the overall increase of the twist of a flare-associated structure and to compare the result to the estimate of the twist based on the field line length and the average force-free parameter at their footpoints.

\begin{figure}
      \centering
      \includegraphics[width=0.9\columnwidth]{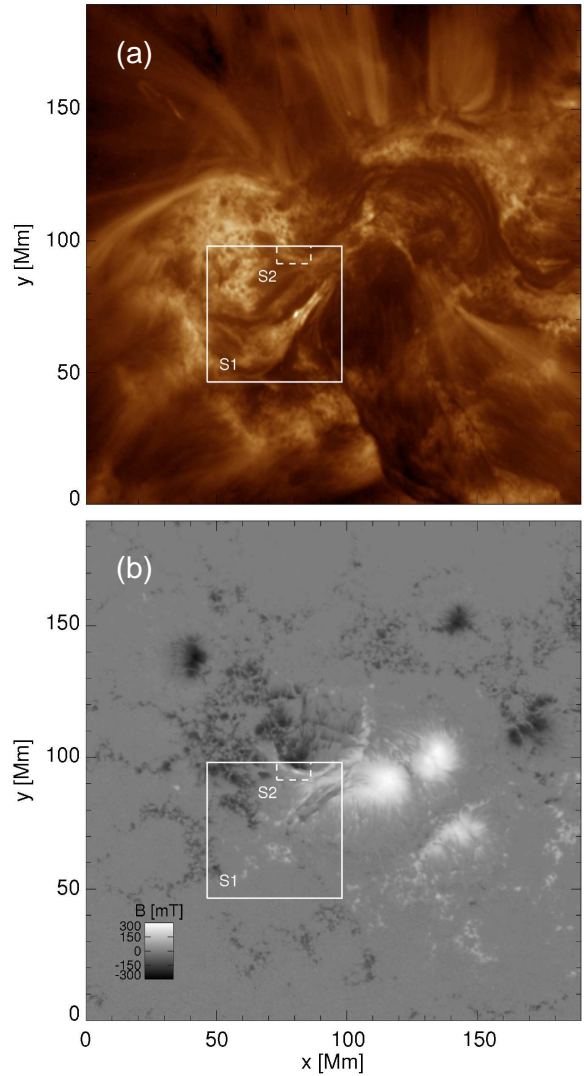}
      \put(-180,385){\large\white\sf(a)}
      \put(-180,190){\large\white\sf(b)}
      \caption{{\bf(a)} AIA 19.3~nm image on 2012 Jul 11 at 19:00~UT, covering a similar region as the Hi-C 19.3~nm observations presented recently by CG13 (compare their Figure~1). {\bf(b)} Vertical magnetic field component of the HMI vector map at 19:00~UT (black/white represents negative/positive polarity). Rectangular boxes outline sub-regions which are used for analysis of a twisted (S1) and a braided (S2) structure. S1 encompasses to a great extend the connectivity of the south-west part of the AR to which a recorded AIA-flare was associated. S2 outlines the region associated to a braided structure, focused on by CG13 (see their Figure~3).}
      \label{fig:fig1}
\end{figure}

\section{Observations and Modeling}

We first align the {\it SDO}/HMI \citep{scho_sche_12} vector maps \citep[][with the 180$^\circ$-ambiguity of the transverse field resolved following \cite{met_94,lek_bar_09}]{bor_tom_10} of NOAA AR 11520 on 2012 Jul 11 at 19:00~UT and a co-temporal AIA 19.3~nm coronal image using standard IDL mapping software. In the same way, we align the Hi-C observation at 18:55~UT and the AIA 19.3~nm observation at 19:00~UT and select sub-fields which cover the field-of-view of the vector maps. The AIA 19.3~nm image (see Figure~\ref{fig:fig1}a) shows patterns of concentrated strong emission (especially above regions of strong negative polarity; compare Figure~\ref{fig:fig1}b) on top of weaker emission on larger scales and weakest emission in the center of the AR where large filament channels run. The strong emission is found especially around $(x,y)$\,$\approx$\,(90,80)~Mm, outlining a narrow, strongly emitting magnetic structure. The Heliophysics Events Knowledgebase\footnote{\url http://www.lmsal.com/hek/index.html} lists an AIA-flare associated to this strongly emitting structure and triggered for being registered by the system in 17.1~nm and 13.1~nm. The small flare started at $\sim$\,18:57~UT and ended at $\sim$\,19:02~UT. 

In absence of coronal magnetic field measurements, NLFF reconstruction techniques based on photospheric magnetic field measurements \citep[within their limitations; see][]{der_schr_09} are to date one of the few means to approximate the coronal field structure with a near real-time temporal cadence, given the spatial resolution provided by the measured field vector  \citep[\eg][]{wie_sak_12}. We take the flux-balanced sub-field of the magnetic vector map (corresponding to the coronal area shown in Figure~\ref{fig:fig1}a) as input for an algorithm to reconstruct the associated NLFF coronal magnetic field. Even though the spatial resolution of the HMI data ($\sim$\,1$''$, and consequently that of the associated NLFF model) is clearly below that of the Hi-C data, we should still be able to grossly estimate the coronal energy content. This is what at best can be done as long as magnetic field measurements of higher resolution (\eg, from the SOT/Spectro-Polarimeter (SP) with $\sim$\,0.6$''$ in fast-mode operation) are not available with a high temporal cadence or during times of flare occurrences (the nearest-in-time SOT/SP measurement was completed about 1~hour before the start of the flare).

HMI vector maps are available at a $\sim$\,12-minute cadence and we search the corresponding sub-fields at 18:00~UT, 18:48~UT and 19:12~UT by means of cross-correlation of the longitudinal magnetic field component. Accounting for projection effects, we transform the magnetic field vectors to the Heliographic coordinate system, \ie~transform the longitudinal and transverse field components to their vertical and horizontal correspondents \citep[following][]{gar_hag_90}. The resulting local magnetic field vectors are then preprocessed following \cite{wie_inh_06} to gain force-free consistent boundary conditions \citep[\eg][]{aly_84,low_85,aly_89} for the NLFF relaxation in the volume above \citep{wie_inh_10,wie_tha_12}. A Laplace problem for the magnetic scalar potential, matching the normal component of the NLFF solution on the volumes' boundary, is solved whose gradient resembles the associated potential field solution.

For the available nearest-in-time SP vector map around 17:54~UT \citep{sku_lit_87,lit_cas_07} we resolve the ambiguity of the transverse field using the same method as used for the available HMI vector products \citep[][]{lek_bar_09}. Hereafter, we treat the obtained vector map in the same way as discussed above for the HMI data and reconstruct the NLFF field above a flux-balanced vector map. We find the sub-region which corresponds to the field-of-view of HMI by cross-correlation of the vertical magnetic field component prior to NLFF modeling. Given the different plate scale of the instruments, this allows us to consider nearly the same sub-volume in the SP model.

From Figure~\ref{fig:fig1}a it is evident that high, over-arching coronal field lines above AR 11520 do not contribute to the AIA emission pattern in the center of the AR. This makes it difficult to verify the NLFF model solution (by comparison of modeled magnetic field lines to coronal loops seen in the AIA image) since only the (open) field at the edges of the active region and some low-lying structures in its center are clearly seen in the AIA image. Much of  the central part of the active region emission is dominated by low-lying dark filament channels. Therefore, we verify the model results by comparison of to the strong AIA emission in the south-east of the AR (in Section~\ref{ss:geometry}). We, however, can indicate the global quality of the NLFF reconstruction based on the HMI vector maps in form of the current-weighted (CW) average of $\sin\theta$, where 0\,$\le$\,$\theta$\,$\le$\,180$^\circ$ is the  angle between the vectors of magnetic field and electric current density \citep[][]{der_schr_09}. We find $\langle\rm{CW}\sin\theta\rangle$\,$\simeq$\,0.07, \ie~$\langle\theta\rangle$\,$\simeq$\,4$^\circ$. (An entirely force-free field gives $\langle\theta\rangle$\,$=$\,0$^\circ$.)

\section{Results}

\subsection{Magnetic free energy of braided structure}

We define ``S2'' (dashed rectangle in Figure~\ref{fig:fig1}), covering the area around the observed braided structure shown in Figure~3b  of CG13. Our NLFF model solutions adhere a spatial resolution of $\sim$\,1$''$ (when based on HMI data) and $\sim$\,0.6$''$ (when basing the modeling on SP data). The braided strands observed by the Hi-C instrument were exhibited angular widths of $\sim$\,0.2$''$, \ie~are below the resolution limit of our models. Nevertheless, we estimate the free magnetic energy, $\Efree$, of the volume around the observed braided structure but assume that the retrieved values represent some lower bound to the real amount of free energy present in the coronal volume. In accordance to the assumption of CG13, we consider a sub-volume of $\sim$\,$10^{11}$~km$^3$ which should cover the observed braided structure and its nearest surrounding (whose ``footprint'' is outlined as S2 in Figure~\ref{fig:fig1}b). From the models based on HMI data (called ``HMI model(s)'' hereafter), we find $\Efree$\,$\propto$\,$10^{23}$~J which is for all evaluated times about 5\% of the total magnetic energy in this sub-volume (see Table~\ref{tab:tab1}).

We repeat the order-of-magnitude estimate as described in CG13 by evaluating the free energy within a certain volume $V$ as to be $\propto$\,$\Bazi^2$\,$V/8\pi$, where $\Bazi$ is the azimuthal field of the braided structure. For the latter, we estimate the magnitude of the average magnetic field in a vertical plane perpendicular to the thought axis of the braided structure. We find $\Bazi$\,$\propto$\,$10^2$~mT and using $V$\,$=$\,$10^{11}$~km$^3$, our analytically estimated amount of free energy becomes $\Efree$\,$\propto$\,$10^{23}$~J. CG13, for comparison, used $\Bazi$\,$\propto$\,10~mT, yielding $\Efree$\,$\propto$~$10^{21}$~J only. Even when taking into account the statistical error of our model-based free energy estimate \citep[$\sim$\,10\% for $\Efree$; see][]{tha_tiw_13}, the estimated free energy we find is much larger than that estimated by CG13 (larger by a factor 10$^2$). Therefore, the discrepancy of the energy estimates may be attributed to the underestimated azimuthal magnetic flux assumed by CG13.

We, additionally compare the free energy estimate at 18:00~UT with that of a NLFF model based on nearest-in-time SP vector map at 17:54~UT (called ``SP model'' hereafter). The order of magnitude agrees ($\Efree$\,$\propto$\,$10^{23}$~J), confirming our free energy estimates from the HMI models. Moreover, we find $\sim$\,60\% more energy in the SP model. This supports the results of a recent case study by \cite{tha_tiw_13} which already indicated that the energy estimates based on models using SP data exceed those of models based on HMI data. 

\begin{table}
      \caption{Magnetic energies associated to field line braiding}
      \vspace{-11pt}
      \begin{center}
            \begin{tabular}{lccc}
                  \tableline
                  ~ & $\Enlff$ & $\Epot$ & $\Efree$ \\
                  ~ & \multicolumn{3}{c}{[$\times$\,$10^{24}$~J]}\\
                  \tableline
				17:54$^\star$ & 3.26 & 2.90 & 0.36\\
				18:00 & 2.10 & 2.00 & 0.10\\
				18:48 & 2.16 & 2.07 & 0.09\\
				19:00 & 2.16 & 2.06 & 0.10\\
				19:12 & 2.15 & 2.06 & 0.09\\
                  \tableline
            \end{tabular}
      \end{center}
      {\scriptsize Total, potential and free magnetic energy ($\Enlff$, $\Epot$ and $\Efree$\,$=$\,$\Enlff$\,$-$\,$\Epot$, respectively) of the 3D model fields in the volume above S2. Non-starred values are based on the HMI models with a resolution of $\approx$\,1$''$ and covering $\sim$\,13.3\,$\times$\,7.0\,$\times$3.7~Mm$^3$, \ie, $\sim$\,3.4\,$\times$\,$10^{11}$~km$^3$.\\
      $^{\star)}$ Values are based on the SP model with a resolution of $\approx$\,0.6$''$ and a volume of $\sim$\,13.3\,$\times$\,6.9\,$\times$3.8~Mm$^3$, \ie, $\sim$\,3.4\,$\times$\,$10^{11}$~km$^3$.\\
       Based on the findings of a previous statistical analysis, the error of the energy estimates can be assumed as $\sim$\,1\% for both $\Epot$ and $\Enlff$ and as $\sim$\,10\% for $\Efree$ \citep[see][]{tha_tiw_13}.}
      \label{tab:tab1}
\end{table}

\begin{figure*}
	\epsscale{1.2} 
      \plotone{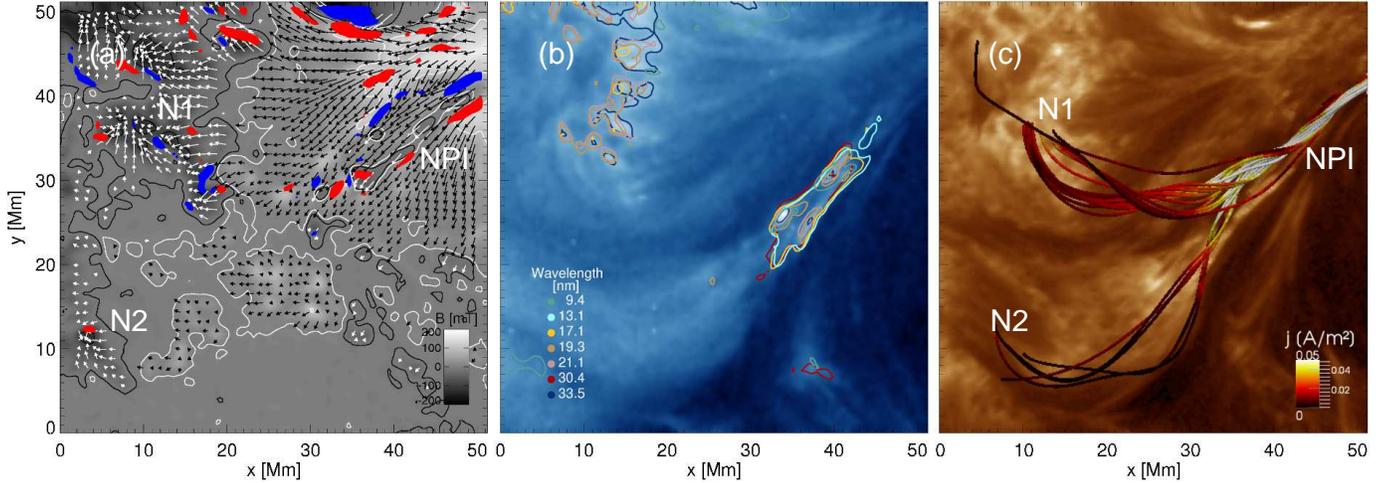}
      \put(-490,160){\large\white\sf(a)}
      \put(-320,160){\large\white\sf(b)}
      \put(-150,160){\large\white\sf(c)}
      \put(-365,120){\large\bf\white\sf NPI}
      \put(-464,139){\large\bf\white\sf N1}
      \put(-482,59){\large\bf\white\sf N2}
      \put(-30,120){\large\bf\white\sf NPI}
      \put(-131,139){\large\bf\white\sf N1}
      \put(-149,59){\large\bf\white\sf N2}
      \caption{{\bf(a)} Sub-field S1 of 2D NLFF lower boundary at 19:00~UT. The gray-scale background reflects the vertical magnetic field, $\Bz$, (black/white represents negative/positive polarity). White/black contours are drawn at $\pm10$~mT. White/black arrows indicate the magnitude and orientation of the horizontal field originating from negative/positive polarity regions where $\Bzabs$\,$>$\,10~mT. A negative polarity island (NPI) is visible as a chain of black closed contours at the north-west of S1. Red/blue filled contours resemble the vertical current density $\jz$ of $\pm$\,0.02~Am$^{-2}$. {\bf(b)} AIA 33.5~nm image covering S1 at 19:00~UT. Contours outline the 98 percentile of the maximum intensity in the 33.5~nm (blue), 30.4~nm (red), 21.1~nm (pink), 19.3~nm (brown), 17.1~nm (yellow), 13.1~nm (cyan) and 9.4~nm (dark green) wavelength channel. {\bf(c)} Selected field lines calculated from the NLFF magnetic field model above S1 at 19:00~UT. The background shows the nearest-in-time Hi-C 19.3~nm observation at 18:55~UT. The field lines are color-coded according to the absolute current density. The view is along the vertical (in negative $z-$) direction.}
      \label{fig:fig2}
\end{figure*}

\begin{figure*}
	\epsscale{1.2} 
	\plotone{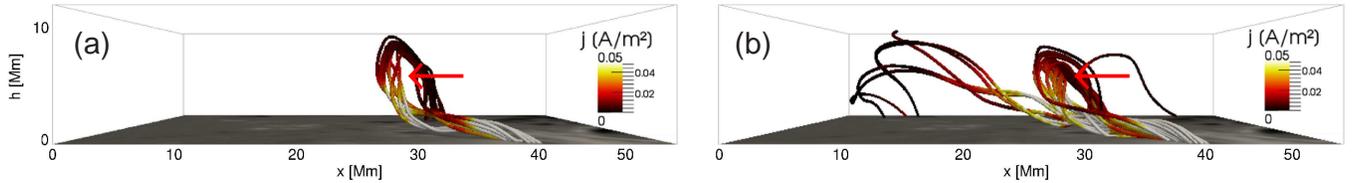}
      \put(-495,50){\large\sf(a)}
      \put(-245,50){\large\sf(b)}
      \put(-118,35){\Huge\bf\red$\leftarrow$}
      \put(-370,35){\Huge\bf\red$\leftarrow$}
      \caption{Selected field lines at {\bf(a)} 18:48~UT and {\bf(b)} 19:00~UT. The field lines in (b) are a subset of the field lines shown in Figure~\ref{fig:fig2}c. The view is along the west-east (negative $x-$) direction. The vertical extension of the box is $\sim$\,12~Mm. The bottom layer reflects the vertical magnetic field component of the NLFF lower boundary.}
      \label{fig:fig3}
\end{figure*}

\subsection{Magnetic field geometry}\label{ss:geometry}

To investigate the magnetic topology associated to the AIA-flare we select a sub-field (solid outline ``S1'' in Figure~\ref{fig:fig1}) which properly adheres the associated magnetic connection. A negative-polarity island (NPI) is discernible in the north-west of S1 (Figure~\ref{fig:fig2}a). The surrounding of the NPI exhibits, besides the regions towards the center of the AR, the highest values of the vertical current density (red and blue filled contours). An inspection of the co-temporal emission in the different AIA channels (Figure~\ref{fig:fig2}b) reveals that the distinct areas of maximum intensity in the various wavelengths are co-spatial. In order to outline brightest structures in the different wavelength channels consistently, the contours outline the 98 percentile for each wavelength channel. This means that the contours outline the region within which pixels of highest intensity are located. Depending on the wavelength channel, either kernels of emission (at 33.5, 21.1, 19.3 and 9.4~nm; presumably outlining substructures) or larger-scale emission (at 30.4, 17.1 and 13.1~nm), surrounding and partly coinciding the former. The fact that the strongest emission in all of the different channels is co-spatial indicates a multi-thermal emission which originates from spatially close regions.

Now, how does the magnetic field configuration which is assumed to be partly outlined by the observed emission look like in detail?  At 19:00~UT, two minutes before the end of the small flare. We assume that the flare-related reconnection in a narrow current sheet somewhere near the strong emission pattern was accomplished by 19:00~UT and that the coronal field was close to a nearly force-free post-flare configuration. As shown in Figure~\ref{fig:fig2}c, a bundle of twisted field lines is present in the reconstructed 3D NLFF field. (Only field lines are shown which start close to that part of the NPI where the positive vertical electric current is strongest; compare Figure~\ref{fig:fig2}a.) The field lines seemingly make up a flux rope which is more compact where it emerges from the lower boundary (in the positive polarity region, bordering the NPI in the north-west of S1) and more extended towards where it re-enters the area of negative polarity in the north-east of S1 (near N1). Comparison to the co-temporal coronal image at 19.3~nm (Figure~\ref{fig:fig2}c) shows that the reconstructed field structure does not perfectly overlap the coronal  emission pattern but does resemble it reasonably well.

Within the flux rope, the strongest values of absolute current density (Figure~\ref{fig:fig2}c) are found at the center and bottom of the tightly twisted parts. When viewed from above, these locations of strong currents well coincide with places of strongest coronal emission, despite a small spatial deviation of a few Mm. This suggests that the observed AIA emission represents dissipated electromagnetic energy which could well be induced by magnetic reconnection in strong electric current concentrations in the twisted flux rope or Joule heating by ohmic dissipation. The repeated brightening in this coronal area observed by CG13, however, supports the former.

\begin{figure*}
	\epsscale{1.2} 
	\plotone{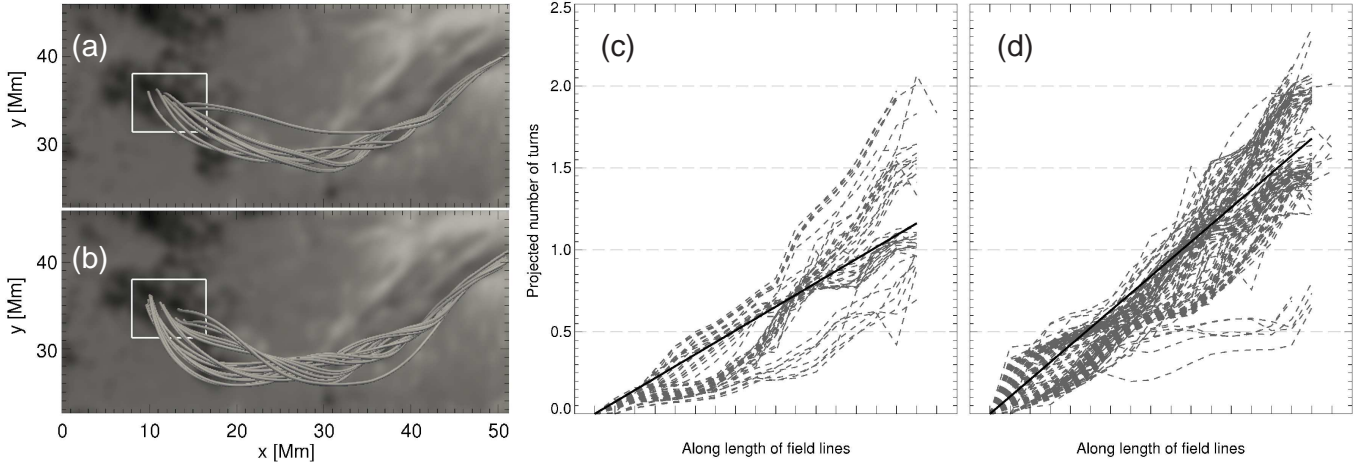}
      \put(-495,155){\large\white\sf(a)}
      \put(-495,75){\large\white\sf(b)}
      \put(-295,155){\large\sf(c)}
      \put(-145,155){\large\sf(d)}
      \caption{Sub-set of all considered field lines, ending in the strongest polarity regions of N1 (white rectangle) on the NLFF lower boundary at {\bf (a)} 18:48~UT and {\bf(b)} at 19:00~UT. The view is along the vertical (negative $z-$) direction. Projected number of turns of each field line pair estimated along a common thought axis at {\bf (c)} 18:48~UT and {\bf(d)} 19:00~UT (gray dashed lines). The black solid line indicates the median gradient of the distribution which represents the overall twist of the entire flux rope.}
      \label{fig:fig4}
\end{figure*}

\subsection{Temporal evolution of the twist}

The field lines of Figure~\ref{fig:fig2}c are again shown in Figure~\ref{fig:fig3}b but when viewed from the side (along the negative $x$-direction). Additionally, we display the corresponding field line geometry at 18:48~UT in Figure~\ref{fig:fig3}a which originate within the same region of strong positive vertical electric current density as those at 19:00~UT. Comparison of the field line geometries 12-minutes apart suggests a reconfiguration of the magnetic field (given the above choice of regarded field lines). While no field lines connect to the negative polarity (N2) in the south-east of S1 at 18:48~UT, some do so at 19:00~UT. This reconfiguration might be caused by magnetic reconnection related to bald batches present at the boundaries of the NPI. Unfortunately, a related investigation is out of the scope of this study.

From a visual inspection of Figure~\ref{fig:fig2}c, the bundle of field lines warps $\sim$\,1.5 times around a thought flux rope axis. From Figure~\ref{fig:fig3} we get the impression that some parts of the flux rope become more twisted with time (indicated by the red horizontal arrows) than other parts and that individual field lines wind more or less often around others. CG13 claimed to see an increasing twist of the structure from the inspection of Hi-C images in the time $\sim$\,18:51~UT -- 18:57~UT. In the following we aim to quantify the overall twist of the flux rope by means of studying the winding of constituent field lines around in the corona about each other. This should provide us with an idea of how much the overall twist increases during the 12 minutes which separate the two model solutions.

We only consider field lines which connect the surrounding of the NPI and the strong negative polarity N1 on NLFF lower boundary (marked by the white box in Figure~\ref{fig:fig4}a and \ref{fig:fig4}b) at 18:48~UT as well as at 19:00~UT. We neglect any field lines which connect the positive polarity around NPI and N2 (from which some are displayed in Figures~\ref{fig:fig2}c and \ref{fig:fig3}b). A subset of all considered field lines is shown in Figure~\ref{fig:fig4}a and \ref{fig:fig4}b which are all field lines originating from the strongest positive vertical current concentration (see Figure~\ref{fig:fig2}a). For each pair of field lines, we calculate their footpoint-to-footpoint winding along their length, \ie~between their line-tied ends. We do so by estimating their relative (projected) position in 3D space (see Appendix for a detailed explanation of the method and the assessment of the uncertainty of the retrieved values). This enables us to determine how often two particular field lines warp around each other (Figure~\ref{fig:fig4}c and \ref{fig:fig4}d for the field lines shown in \ref{fig:fig4}a and \ref{fig:fig4}b, respectively). We then assume that the median number of turns of all possible field line pairs within the flux rope is representative for its {\it overall} twist (represented by the black solid line in Figure~\ref{fig:fig4}c for 18:48~UT and \ref{fig:fig4}d for 19:00~UT). The term {\it overall} is also to account for the fact that our method does not distinguish between the twist of the flux tube axis itself and the winding of the field lines with respect to that twisted axis. The changing steepness of the distributions of the estimated overall twist immediately suggests an increase from $\sim$\,1 to $\sim$\,1.5~turns.

We chose the above discussed subset of field lines ($\sim$\,10 when using a spatial sampling of 0.5$''$ and allow the field lines to originate only from areas of strong vertical electric current), to be able to show clearly represented graphs in Figure~\ref{fig:fig4}c and \ref{fig:fig4}d, where we display the winding of every possibly combination of pairs of field lines ($\sim$\,50). For calculation of the overall twist, however, we use a much larger number of field lines ($\sim$\,10$^2$ with a finer spatial sampling of 0.25$''$ and by allowing the field lines to originate anywhere near the NPI, yielding $\sim$\,500 field line pairs to be considered). Additionally, we vary the area around the NPI from which field lines have to connect to N1 as well as in- and decrease the spatial sampling of field line footpoints. This  allows us to estimate the uncertainty of our overall twist estimate in terms of field line selection. This analysis yields an overall (median) twist $\widetilde{T}$\,$=$\,1.2\,$\pm$\,0.03 turns at 18:48~UT and $\widetilde{T}$\,$=$\,1.7\,$\pm$\,0.05 turns at 19:00~UT. The given uncertainty represents the mean absolute deviation from the median, defined as the mean of the absolute deviations from the median itself ($\sum(|x_i-\widetilde{x}|)/N$, where $x_i$ are the elements of a sample, $\widetilde{x}$ its median and $N$ is the number of elements). This means that, on overall, the field line configuration acquires more twist in the course of the AIA flare (about half a turn within 12 minutes). This not only quantitatively confirms what was suspected from the visual inspection of the NLFF model field line configurations (Figure~\ref{fig:fig3}), it also verifies what was suspected by CG13 through a pure visual analysis of a 5-minute sequence of coronal images, namely that the twist in the flare-related structure increased with time. A corresponding analysis of the magnetic field configuration at 19:12~UT reveals an ongoing but less rapid increase of the overall twist. 

For comparison, we also estimate the twist of the individual field lines using a method as often found in literature \citep[\eg,][]{lea_can_03,ino_shi_12}. For each field line of the considered subset we calculate $T_{\overline{\alpha}}$\,=\,$\overline{\alpha}L/4\pi$ (where $\overline\alpha$ is the mean value of the force-free parameter $\alpha$\,$=$\,$\mu_0\,\jz/\Bz$ at both footpoints and $L$ is the arc-length of the field line). Here, we find a median of $\widetilde{T}_{\overline{\alpha}}$\,$=$\,0.7\,$\pm$\,0.03 at 18:48~UT and $\widetilde{T}_{\overline{\alpha}}$\,$=$\,1.1\,$\pm$\,0.03 at 19:00~UT. Firstly, this result supports the estimated overall twist increase based on our purely geometrical analysis ($\sim$\,0.5~turns within 12 minutes). Secondly, the fact that we find lower values for the overall twist from the latter method underlines what was argued by \cite{lek_fan_05}, namely that force-free $\alpha$ based methods may underestimate the twist of larger-scale structures within ARs. The increase of the overall twist of the 3D magnetic structure is naturally related to a correspondent change in the underlying magnetic field. The median value of the force-free parameter of all analyzed field lines increases from $\widetilde{\alpha}$\,=\,0.4\,$\pm$\,0.02~Mm$^{-1}$ at 18:48~UT to 0.6\,$\pm$\,0.08~Mm$^{-1}$ at 19:00~UT.

To judge the influence of spatial resolution on our geometrical twist estimate, we additionally compare the overall twist of the configuration in the SP model at 17:54~UT with that of the nearest-in-time HMI model (at 18:00~UT). The resulting overall twist of $\widetilde{T}$\,$=$\,1.0\,$\pm$\,0.2 turns at 17:54~UT (based on the SP model with a resolution of $\sim$\,0.6$''$) and $\widetilde{T}$\,$=$\,1.2\,$\pm$\,0.0 turns at 18:00~UT (based on the HMI model with a resolution of $\sim$\,1$''$) agrees within the error ranges (which represent again the mean absolute deviation from the median). We find a similar median value of the force-free parameter from both, the higher-resolution SP ($\widetilde{\alpha}$\,=\,0.5\,$\pm$\,0.08~Mm$^{-1}$) and lower-resolution HMI ($\widetilde{\alpha}$\,=\,0.5\,$\pm$\,0.1~Mm$^{-1}$) model.

\section{Summary and Discussion}

Just recently, through the observation of the coronal plasma with unprecedented spatial resolution, new insights on the processes heating the solar corona were gained. Most plausibly, such processes involve the reconfiguration of the magnetic field at coronal heights since the associated magnetic energy outclasses the kinetic, thermal and gravitational energy \citep[\eg][]{for_00}. Following an analytical expression to number the free magnetic energy associated to a spatially resolved bundle of braided coronal loops \citep[given by][hereafter ``CG13'']{cir_gol_13} an amount of $\propto$\,$10^{21}$~J of free magnetic energy was suspected. Furthermore, they estimated that about 0.1\% of the stored energy was converted into the observed radiation. Moreover, they interpreted the analyzed sequence of coronal images as to depict an increase of the twist of a magnetic structure during the rising phase of a small flare.

In this study, we aimed to verify the findings of CG13 by compensating the deficiency of direct observations of the coronal magnetic field by reconstructing the associated AR coronal nonlinear force-free field. We used measurements of the photospheric field vector by {\it SDO}/HMI and {\it Hinode}/SOT-SP (with a plate scale of $\sim$\,0.5$''$ and $\sim$\,0.3$''$, respectively) and employed the associated, static nonlinear force-free equilibrium solution in the 3D model volume above. We first checked the free energy within a volume of $\sim$\,$10^{11}$~km$^3$ containing the braided structure and estimated an amount of  $\propto$\,$10^{23}$~J, \ie~$\sim$\,10$^2$ times more than what the analytical estimate of CG13 delivered. We were able to attribute this difference to the underestimated strength of the azimuthal magnetic field of the observed braided structure. This firstly, highlights the importance of the analysis of the coronal magnetic field energy based on vector magnetic field measurements and/or force-free model techniques and secondly, allows us to conclude that even more free magnetic energy is available for heating the AR corona than what was suspected from the observational analysis of the AR corona.

We furthermore investigated a magnetic flux rope which has been associated to a small AIA-flare. Strongest absolute current density was found in those parts of the flux rope which were most tightly twisted. We were able to associate the highly twisted parts of the flux rope to highest coronal emission in all AIA wavelength channels. We interpret this spatial overlap by the field-aligned currents being dissipated in the course of magnetic reconnection or ohmic dissipation and yielding the radiative losses which are observed in form of coronal emission signatures.

We also looked at the temporal evolution of the {\it overall} twist of the flux rope by directly incorporating the modeled 3D geometry of the constituent magnetic field lines. To our knowledge this is the first time that the shape of modeled field lines has been used to quantify the average twist of a flux rope. For each pair of a subset of field lines making up the twisted flux rope we estimated their footpoint-to-footpoint winding. The median number of turns of all possible combinations of field line pairs allowed us to derive the increase of the overall twist of the flux rope (from about 1.0 to 1.7~turns within $\sim$\,12 minutes).  Additionally, we used a method as commonly used in literature to calculate the AR-twist which involves the force-free parameter at the line-tied field line footpoints and the field line length. Here we found similar result, namely that the average winding of the flux rope increased by about half a turn in the course of the small flare. This allowed us to confirm the assumption of \cite{lek_fan_05}, namely the ability of the latter method to adequately recover the winding of thin flux tubes (which we assume our field lines are) but to slightly underestimate the twist of larger-scale structures. For investigating the effect of spacial resolution, we also employed the overall twist of the structure at around one hour before the flare based on a higher-resolution SP model and a nearest-in-time ($\sim$\,6~minutes apart) lower-resolution HMI model. We found a similar overall twist from both models agreeing to within the statistical error. The overall twist found from the earlier-in-time SP model was slightly lower than that found from the HMI model which could also be due to the temporal evolution of the magnetic field during the 6 minutes separating the models.

In summary, using a sequence of nonlinear force-free coronal magnetic field models, we (1) were able to confirm the ability of the delivery of magnetic energy by braided magnetic field lines sufficient for the heating of the solar corona, (2) were capable to associate the localized strong coronal emission and the strong localized field-aligned currents in a twisted flux rope and (3) presented a novel approach to quantify the temporal evolution of the overall twist of a flare-related structure. Our investigation, on the one hand, supports the conclusion drawn by CG13: the free magnetic energy stored in the low-lying coronal loops (braided magnetic field lines) of an AR is sufficient to heat the AR corona by radiating the heat which is delivered by small-scale flares. On the other hand, our work underlines the great potential of force-free coronal field models to partially explain observational emission signatures by the associated modeled coronal magnetic field structure, which at present time is not routinely accessible via direct measurements.

\acknowledgments

We thank the anonymous referee for careful consideration on this manuscript and useful comments. J.\,K.\,T., acknowledges support from Austrian Science Fund (FWF): P25383-N27 and DFG grant WI 3211/2-1. T.\,W.~is funded by DLR grant 50 OC 0904. {\it SDO} data are courtesy of the NASA/{\it SDO} AIA and HMI science teams. {\it Hinode} is a Japanese mission developed and launched by ISAS/JAXA, with NAOJ as domestic partner and NASA and STFC (UK) as international partners. It is operated by these agencies in co-operation with ESA and NSC (Norway).  {\it Hinode} SP Inversions were conducted at NCAR under the framework of the Community Spectro-polarimetric Analysis Center. We acknowledge the Hi-C instrument team for making the flight data publicly available. MSFC/NASA led the mission and partners include the SAO in Cambridge (MA); LMSAL in Palo Alto (CA); UCLAN in Lancashire (UK); and the LPI RAS in Moscow.

\appendix

\section{Estimation of field line twist}

To estimate the overall twist of the flare-related flux rope, we consider each pair of all considered field lines separately (red and black solid line in Figure~\ref{fig:fig5}a). The average footpoint position of each field line pair is connected by a thought ``principal axis'' (PA) of a thought thin flux tube thought to be composed of the two field lines. Planes normal to the PA (``axis-normal (AN) planes'') are defined and used for further analysis (in this particular case 19 planes; dashed colored lines in Figure~\ref{fig:fig5}a). We determine the locations in 3D space where both of the field lines intersect these AN planes and project them into a common plane (colored diamonds in Figure~\ref{fig:fig5}b, the color accords to the respective AN plane in Figure~\ref{fig:fig5}a which a field line intersects). This allows us to calculate the angle spanned between each of the pair of intersections (diamonds linked by dotted lines of same color; small panel in Figure~\ref{fig:fig5}b; each angle is color-coded according to the dotted lines connecting pairs of intersections). The spanned angle naturally is in the range $-\pi\le\theta\le\pi$ and we take the angle measured from the first intersection as reference (zero) angle. Thus we are able to count how often two field lines are winding around each other along the thought common axis. For the example field line pair discussed here about 1.5 turns (gray dotted line in Figure~\ref{fig:fig5}c). The average gradient of the winding curve of each field line pair delivers the average twist of the thought thin flux tube they define (black dashed line). The median winding of all possible pairs of field lines which connect the NPI and N1 in Figure~\ref{fig:fig4}a finally delivers the overall twist of the entire flux tube.

\section{Applicability and uncertainty of the method}

The ability of our method to reasonably recover the twist of  a flux rope depends also on its thickness. The thinner it is (\ie~the smaller its cross section compared to, \eg, its length or curvature radius) the better our method is expected to work. These are flux ropes whose constituent field lines are everywhere close in space, \ie~where the common thought flux rope axis and consequently the PA well represent each of its constituent field lines. However, there might be pairs of field lines whose line-tied ends are close by each other on one end of the field lines but whose footpoints on the other end of the field lines are located far away from each other (like in case of, \eg, an expanding flux tube). The present analysis compensates partly for this for that we considered only field lines which connect the NPI and N1, \ie~by selection the field lines' footpoints should be relatively close in space. We find that the mean flux tube diameters ($\overline{d}$; determined by the mean relative footpoint positions of a constituent pair of field lines) are by a factor 10 -- 10$^3$ smaller than the mean flux tube length ($\overline{l}$; defined by the mean arc-length of the field line pair). About 90\% (10\%) of the considered field line pairs exhibit $\overline{d}/\overline{l}$\,$\le$\,0.05 (0.01). Therefore, we assume that the thin flux tube approximation holds for most of the considered pairs of field lines.

In Section 3.3 we already gave an uncertainty for the estimated overall twist of the flux rope, depending on the specific choice of field lines used for analysis (in terms of restricting the consideration of field lines to certain connectivity domains and/or spatial sampling). Here, we also note the influence of the number of AN planes used for the analysis (19 in the presented case). We find that the results for the overall twist are almost identical when using $\sim$\,10 or $\sim$\,10$^3$ AN planes (or anything in between) and that the uncertainty conforms with that given in Section 3.3, namely $\propto$\,0.1~turns.

\begin{figure}
      \centering
      \includegraphics[width=\columnwidth]{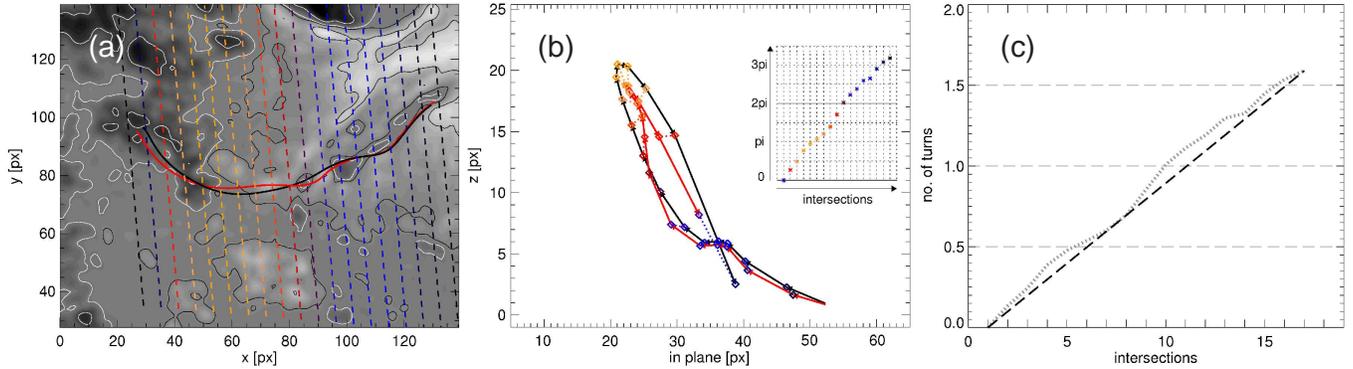}
      \put(-480,120){\large\white\sf(a)}
      \put(-310,120){\large\sf(b)}
      \put(-135,120){\large\sf(c)}
      \caption{{\bf(a)} A pair of randomly selected field lines from the subset of field lines shown in Figure~\ref{fig:fig4}b at 19:00~UT. A principal axis (PA) is defined by the mean start and end location of the footpoint locations of the field line pair. Dashed lines mark planes perpendicular to and along the PA (called ``axis-normal'' (AN) planes). {\bf(b)} Intersections of the field lines with the AN planes, projected into the a common plane. Colored diamonds correspond to projected intersections of the field lines with the AN planes equally colored in (a). The small figure in (b) shows the progression of the angle calculated between each pair of field line intersections with respect to the first measured (``zero'') angle. {\bf(c)} Resulting winding number of the field line pair (gray dotted curve) as estimated from the projected angles in (b). The black dashed line shows the median value of the gradient, corresponding to the average winding of the particular field line pair.}
      \label{fig:fig5}
\end{figure}

Another uncertainty arises due to the fact that the AN planes (dashed lines in Figure~\ref{fig:fig5}a) are everywhere normal to the PA (joining the mean footpoint position of a considered field line pair) but might not be normal to the the actual flux tube axis all along the length of it. Some of the planes might be at an oblique angle with respect to the actual flux tube axis (especially towards the ends of the line-tied field lines; see Figure~\ref{fig:fig5}a), in extreme cases even parallel. And the flux tube axis itself might be twisted which is not taken account for in the presented method. Consequently, the relative (projected) distance of different portions of the considered field lines in space (Figure~\ref{fig:fig5}b) are afflicted with a greater or lesser uncertainty. To test our results of the overall twist of the entire ensemble of considered field lines, we tilt the AN planes with respect to the PA-normal direction until they are almost parallel to the PA (\ie, we tilt them up to $\pm$\,80$^\circ$ with respect to the PA-normal direction). The arising uncertainty for the overall twist conforms with the uncertainty ranges discussed above.

\bibliographystyle{apj}
\bibliography{jkt_apj_a}
\end{document}